# What do you mean by 'human–AI collaboration'?
## Prerequisite functions and the affordances needed to achieve it

**Professor Mutlu Cukurova**

**UCL Knowledge Lab & UCL Centre for Artificial Intelligence,**

**University College London, London, UK**

**Email:** m.cukurova@ucl.ac.uk

ORCID: 0000-0001-5843-4854

## Abstract

The concept of 'collaboration' has been extended rapidly to describe what people now do with conversational agents, intelligent tutors, adaptive platforms, and generative artificial intelligence (AI) tools in general. This chapter asks what is gained and lost when a demanding concept from the learning sciences is applied so freely. Returning to long-standing accounts of collaborative learning, it reconstructs the requirements that a situation, an interaction, and a set of cognitive processes have historically had to meet before being called collaborative. Human-AI collaboration requires a partly symmetric and negotiated relationship, shared and negotiable goals, a low and shifting division of labour, interactive and synchronous exchange, and mutual modelling, grounding, and socially shared regulation. Reviewing process-sensitive empirical studies of writing and problem solving, the chapter shows that most current human–AI interaction is better described as consultation, governance, delegation, or instruction rather than as collaboration. To make these distinctions functional, the chapter introduces a five-level diagnostic taxonomy of human–AI teaming (i.e. transactional, situational, operational, praxical, and synergistic) defined by the affordances an AI system exhibits. It shows that only the highest level begins to satisfy the conditions the tradition places on collaboration. The chapter derives the functions an AI system must possess for collaboration to be achievable, argues that most of these are present-day engineering choices rather than capabilities to be awaited, and sets out the implications for research, measurement, and responsible practice of human-AI collaboration in education.

# 1. Introduction

A quarter of a century ago, Dillenbourg (1999) opened a now-canonical chapter with a frank admission; a multidisciplinary task force convened to study collaborative learning had failed to converge on a definition of its central object. The reason was not failure of imagination but proliferation of usage. ‘Collaboration’ had become a fashionable term back then, applied, in Dillenbourg’s phrase, “abusively for more or less anything”. That promiscuity carried two costs. The first was theoretical. Claims about the cognitive effects of collaboration become vacuous once any human-human interaction can be called collaborative. The second was scholarly. Contributions from psychology, education, and the computational sciences become difficult to articulate when the same word is used in incommensurable ways across disciplines.

The present chapter resurfaces Dillenbourg’s question for the era of large language models, co-pilots, and agentic AI systems. The label ‘collaboration’ has been borrowed once more. This time to describe what humans now do with conversational agents, intelligent tutors, adaptive learning platforms, and a rapidly proliferating ecology of generative AI tools. Reviews of human–AI teaming (O’Neill et al. 2022; Berretta et al. 2023; Marrone et al. 2025; Zamecnik *et al.* 2026) and of artificial intelligence in education (Cukurova 2025a; Holstein et al. 2020; Cukurova et al. 2026) report persistent terminological inconsistency, while public discourse routinely personifies systems whose actual properties they do not yet possess (Salles et al. 2020; Shanahan 2024). The risk Dillenbourg flagged in 1999 has therefore returned in an amplified form that if every useful interaction with an AI system is labelled human-AI ‘collaboration’, the construct ceases to discriminate among genuinely different mechanisms, and the empirical and design programmes that follow from it lose their theoretical and empirical grounding.

The stakes of this terminological confusion are not merely scholarly. Educational policy increasingly invokes human–AI ‘collaboration’ as both an aspiration and a curriculum target, and design briefs for new educational technologies typically promise ‘collaborative AI partners’ as if the construct had a stable definition and referent. Where it does not, the cost is paid at three layers. First, the science loses the construct’s discriminative power; second the engineering loses its design targets; and third learners and teachers acquire expectations that the systems they encounter cannot fulfil. The argument for conceptual clarity is thus also an argument for responsible practice.

My purpose in this chapter is four-fold. First, I revisit the requirements that the learning sciences have historically placed on a situation, an interaction, and a set of processes before calling them collaborative, and I argue that these requirements should remain a high bar. Second, I bring in empirical evidence on how human–human and human–AI interactions differ in the same interaction contexts, drawing on process-rich studies of writing and problem solving. Third, I introduce a five-level diagnostic taxonomy of human–AI teaming (i.e. transactional, situational, operational, praxical, and synergistic; Cukurova et al. 2026) which distinguishes qualitatively different human–AI interaction types and names the functions each type presupposes, and I argue that only the topmost level even attempts to satisfy the demands the learning-sciences tradition places on collaboration. Fourth, I develop a functional ontology of collaboration and derive from it the affordances an AI system must possess for collaboration to be achievable at all. At last, I close the chapter with a discussion of what follows for the field's research agenda.

The argument that follows is therefore neither dismissive nor optimistic. It is an attempt to keep the concept of 'collaboration' doing what it was meant to do which is to name a particular, demanding mode of joint cognition whose presence we should be willing to assert only when its underlying conditions are met.

## 2. The variety of meanings for 'human–AI collaboration'

Dillenbourg (1999) disaggregated 'collaborative learning' into four loci (i.e the situation, the interaction, the cognitive processes, and the effects) and then asked, for each locus, when the modifier collaborative was earned. The same disaggregation can be implemented as the analytical lens for human–AI interaction.

In the current literature, the situation has been called collaborative in a diverse set of interaction including when a teacher and an AI tutoring system share oversight of a classroom (Holstein et al. 2018), when a writer and a generative model produce a text together (Lee et al. 2022), and when a teacher consults a learning analytics dashboard (Wise and Jung 2019). Interactions have been called collaborative when learners prompt a chatbot (Kasneci et al. 2023), when an intelligent tutoring system adapts its sequencing to a student's responses (Aleven et al. 2023), and when a teacher reorganises groups in light of a system's recommendation (Holstein et al. 2019). Processes have been called collaborative when learners regulate their own effort in dialogue with an AI tutor (Molenaar 2022) and when human writers internalise lexical choices initially proposed by a model. Effects have been called collaborative when post-tests improve, when efficiency rises, or simply when satisfaction scores are favourable (Bauer et al. 2025).

These are not equivalent claims, and several of them stretch 'collaborative' well beyond what the learning-sciences literature has ever asked of it. The most parsimonious diagnosis is that the modifier is doing rhetorical rather than analytic work. It signals warmth and equality where the underlying relation is one of instruction, retrieval, or delegation. One could hardly imagine an analogous expansion in adjacent fields. For instance, we do not speak of human–calculator collaboration, despite calculators being responsive, accurate, and persistently useful. Where the modifier is doing genuine analytic work, by contrast, it should be possible (similar to human-human collaboration) to point to specific features of the situation, of the interaction, of the cognitive processes, and of the resulting outcomes.

The slippage is compounded by the proliferation of adjacent constructs, each of which names something specific and none of which is interchangeable with 'collaboration'. Hybrid intelligence names the socio-technical ensemble in which human and artificial information processing are coupled so that the ensemble achieves what neither could alone (Dellermann et al. 2019; Cukurova 2025a). ‘Machines as teammates’ names a research programme on autonomous, proactive technologies participating in team decision-making (Seeber et al. 2020). Human–autonomy teaming names the human-factors tradition concerned with interdependence and coordination between humans and autonomous agents (O’Neill et al. 2022). Mutual theory of mind names the reciprocal construction and revision of models of the other party in communication (Wang and Goel 2022). Each of these constructs has its own conditions of application, and the casual substitution of ‘collaboration’ for any of them erases distinctions the field has worked hard to draw.

It is worth pausing on the analytic costs of the slippage. A first cost is empirical. When ‘collaboration’ ranges over everything from chatbot prompting to teacher–dashboard consultation, studies that purport to evaluate ‘the effects of human–AI collaboration’ aggregate heterogeneous mechanisms whose effects ought to be measured separately (see for details of a similar arguments Cukurova *et al*. 2018). A second cost is methodological. The field acquires a strong incentive to measure surface markers (e.g. satisfaction, output quality, perceived helpfulness) because the underlying construct supplies no purchase for deeper process measures. For instance, Gomez et al.’s (2025) systematic review of 105 empirical studies of AI-assisted decision-making reached a conclusion its own title announces ‘human–AI collaboration is not very collaborative yet’ finding the observed interactions dominated by simplistic accept-or-reject paradigms with little support for genuinely interactive functionality. A third cost is conceptual. The term begins to crowd out the more precise vocabulary the field already possesses (e.g. cooperation, coordination, instruction, augmentation, orchestration), each of which captures something specific that ‘collaboration’ then mis-names (Similar to Berretta et al.’s (2023) scoping review of the human–AI teaming literature which found no consistent definition or terminology across the corpus, which remains dominated by technology-centric rather than team-oriented perspectives).

## 3. Requirements for collaboration

Three traditions converge on a recognisable, if not unitary, account of what collaboration requires. The socio-cognitive tradition (Doise and Mugny 1984; Perret-Clermont et al. 1991) emphasises symmetry of action, knowledge, and status, and the conflicts and negotiations such symmetry enables. The socio-cultural tradition (Rogoff 1990; Vygotsky 1978; Wertsch 1985) emphasises internalisation and the appropriation of tools, signs, and partners’ utterances. The shared-knowledge tradition (perhaps best described by Roschelle and Teasley’s (1995, p. 70) that collaboration is ‘a coordinated, synchronous activity that is the result of a continued attempt to construct and maintain a shared conception of a problem’) emphasises grounding, mutual modelling, and the maintenance of a Joint Problem Space. Drawing on these traditions, I argue (Cukurova 2025a) that at least five requirements characterise collaboration in a strong sense. This list is not exhaustive, but it identifies the conditions that recur, in differently weighted forms, across the canonical reviews and research traditions.

## 3.1 A negotiated and partly symmetric situation

The first requirement is that the situation permit some non-trivial degree of symmetry. Dillenbourg (1999) is careful to distinguish symmetry of action, of knowledge, and of status. Pure symmetry is rare and not strictly necessary for collaboration; what matters is that asymmetry remains dynamic, negotiated, and mutually consequential. Where one agent's role is fixed by design (for instance a tutor cannot be challenged; or a student cannot set the agenda or goals for learning) the situation slides toward instruction, coaching, or coordination. These are valuable working relationships in their own right; but none of them is collaboration. In human–AI cases, symmetry is overwhelmingly constrained at design time; the user prompts and the system responds, the user accepts or rejects, and the action space available to each party is asymmetric not because of an emergent division of labour but because the technical architecture often admits only one direction of initiative.

## 3.2 Shared and negotiated goals

The second requirement is that the agents pursue, and recognise themselves as pursuing, a shared goal whose precise content is at least partly the object of negotiation. Dillenbourg illustrates this point with his lost-keys thought experiment; a thief who takes advantage of my distraction is not my collaborator, because I never intended to help. Coincident outcomes do not amount to shared intentions. Bratman's (1992) analysis of shared cooperative activity, developed in parallel, makes the same point philosophically; the agents must each intend that they perform the action together, and each must adopt subplans that mesh with the other's. Developmental research gives this requirement its empirical teeth which is often named as the capacity for shared intentionality (the motivation and ability to share psychological states and jointly commit to goals) emerges early in human ontogeny, is scarcely present in our nearest primate relatives, and is plausibly the cognitive foundation on which all subsequent collaborative cognition rests (Tomasello et al. 2005). The Roschelle and Teasley (1995)'s construct of the 'Joint Problem Space' also operationalises this requirement. It is precisely the negotiated conception of the task that partners build and rebuild as they work. Current AI systems can be given goals; yet it is difficult to say that they intend goals.

## 3.3 Low and shifting division of labour

The third requirement, on which Dillenbourg insists particularly strongly, is that labour not be cleanly partitioned at the outset. Similar to earlier conceptualisations (see for instance Damon and Phelps 1989) cooperation, in his usage, is the term reserved for that arrangement; split the task, work in parallel, assemble the parts. Collaboration involves a horizontal rather than vertical division, in which one partner can take responsibility for monitoring while the other carries out, and these roles can flip within minutes. Stable specialisation (similar to the kind found in command-and-execute relations) is the structural opposite of collaboration. Miyake's (1986) constructive-interaction studies remain the locus classicus for this type of distinction. The productive engine of paired problem solving is precisely the unstable horizontal partition between executor and monitor. Most human–AI workflows, by contrast, freeze the horizontal partition at the outset and, in most cases, the human regulates and verifies, the system generates, and the roles do not interchange.

## 3.4 Interactivity, synchronicity, and negotiability

The fourth requirement concerns the interactions themselves. Collaborative interactions are interactive in a strong sense that is the interleaving of utterance and action is so tight that interaction continuously reshapes reasoning. They are subjectively synchronous, in the sense that each agent can act on the working assumption that the other is reasoning along with it (Dillenbourg and Traum 1996). And they are negotiable. Positions are offered as positions, not pronouncements, and meta-communication about how to interact (for instance about whether 'that was a question or a claim', about who has the floor etc.) is itself available as a resource (Baker 2015). It is worth noting how thoroughly current AI interaction often breaches these conditions. In most human-AI interaction, turn-taking is rigidly serialised, the system rarely interrupts even when correction is warranted, and meta-communicative repair is almost entirely the user's burden (Cukurova 2025a).

## 3.5 Mutual modelling, grounding, and socially shared regulation

The fifth requirement is potentially the deepest and the one most often missed. Collaboration requires that each agent maintain a working model of the other (partial, revisable, and inferentially active) and that the two models be brought into approximate alignment through grounding (Clark and Brennan 1991). The cognitive science of joint action shows how deep this runs in human dyads and partners co-represent one another's actions and predict their timing involuntarily, even when doing so degrades their individual performance (Sebanz et al. 2006). Grounding is not a polite preamble to collaboration; it is the mechanism by which the 'Joint Problem Space' is built and repaired (Roschelle and Teasley 1995). And because cognitive load is distributed across the dyad, collaborative dyads typically develop a socially shared regulation of learning (i.e. joint monitoring, evaluation, and adjustment of their progress) that exceeds anything either agent could mount alone (Hadwin et al. 2018; Järvelä et al. 2021). Recent work has begun to ask whether AI can participate in such regulation directly, as a co-regulating partner rather than a mere scaffold (Järvelä et al. 2023; Molenaar 2022); but answers are not very clear yet.

This requirement, more than any of the others, reveals why mere conversational fluency is not collaboration. A system can produce utterances that look like back-channelling, reformulation, or even challenge without doing any of the underlying work that those utterances would index in a human partner. Wertsch's (1985) reanalysis of Vygotsky locates the developmental motor of internalisation not in the transition from interpsychological to intrapsychological functioning but in the moment at which the child becomes able to use, in the joint plane, the very tool that will later be internalised. That moment requires a partner who not only models the learner but is changed by the learner.

Current systems can do the former, in a limited way (Hooshyar *et al*. 2026); they cannot, durably, do the latter.

Taken together, these five requirements characterise collaboration as a graded construct, not a binary one. What they make jointly clear is that collaboration is not simply the presence of another agent. It is what happens when joint activity is sustained at a high level along several of these dimensions at once.

# 4. Interaction differences: humans with humans versus humans with AI

If on agrees that those are the requirements, then the question becomes empirical. In the situations now widely described as human–AI collaboration, are the interactions actually collaborative, and if not, in what specific ways do they differ? Four findings recur across the most recent process-sensitive studies.

## 4.1 Distributions of use are heterogeneous, and the modal pattern is extractive rather than collaborative

When learners are given open access to generative AI in authentic tasks, their use does not converge on anything recognisable as joint problem solving. For instance, Stojanov et al.'s (2024) latent profile analysis of 490 university students' self-reported reliance on ChatGPT identified five distinct profiles i) versatile low reliers (38.2%), ii) knowledge seekers (16.5%), iii) proactive learners (11.8%), iv) assignment delegators (23.1%), and v) all-rounders (10.4%). Of these, only the proactive learners (roughly one student in eight) exhibit the feedback-seeking, planning, and self-quizzing uses that has the potential to be a collaborative dialogue; the assignment delegators, nearly twice as numerous, rely on the system precisely to produce their work for them. But, the authors didn't investigate process-traces of student interactions to further investigate if this category's interaction is indeed collaborative.  Similarly, analysing 626 logged interaction activities of doctoral students writing with a GPT-based assistant, Nguyen et al. (2024) found that students who engaged the system iteratively, treating outputs as objects to interrogate and revise, achieved higher writing performance, while those who used it as a supplementary information source within a linear 'prompt–copy–paste' workflow performed worse, the AI's output functioning as placeholder text rather than as a trigger for conceptual development. By contrast, in studies of human–human collaboration on equivalent tasks (Roschelle 1992; Schwartz 1995), iterative co-construction is the rule rather than the exception, because the partner's contribution itself demands uptake. With AI, uptake is optional, and most learners decline.

The costs of declining are now experimentally documented. In a randomised controlled trial with nearly a thousand high-school mathematics students, unrestricted access to a GPT-4 interface raised assisted practice performance by 48 per cent; yet when access was subsequently withdrawn, the same students performed 17 per cent worse than controls who had never used the tool (Bastani et al. 2025). Convergent experimental evidence shows that LLM use lowers learners' cognitive load relative to traditional web search, but at the price of measurably shallower reasoning in the resulting analyses (Stadler et al. 2024). In the vocabulary of the ICAP framework (Chi and Wylie 2014), the modal human–AI exchange hovers at the passive–active boundary; the constructive and interactive engagement that most strongly predicts learning must be imported by the user, against the grain of the design.

This is more than a behavioural observation; as it has a structural cause. Human partners impose what we might call interactional pressure, which is the obligation to respond, to clarify, to take a position, that an AI partner does not, because the AI carries no analogous obligation in the other direction. The cost of leaving an AI's output unengaged is borne by the user alone. In human–human dyads, by contrast, mis-uptake produces visible interactional damage (e.g. the partner repeats, escalates, withdraws), which functions as a continuous incentive toward genuine co-construction.

## 4.2 Grounding and mutual modelling are unidirectional or absent

The clearest qualitative difference between human–human and human–AI interactions concerns mutual modelling. In dyads, partners continually infer what the other knows, intends, and is about to do, and they update those inferences from microscopic cues (Clark 1996; Pickering and Garrod 2004). With current AI, no such update occurs on the system side beyond the duration of a single context window. The system has no durable model of the user across sessions; no representation of what this user, in particular, has come to believe, doubt, or commit to; and no representation of itself as having committed to a position. Users, for their part, can construct elaborate models of the AI (Glikson and Woolley 2020), but the asymmetry is structural as the model is not reciprocated, only simulated. Even in highly fluent co-writing sessions, the human does almost all the partner-modelling work, while the system's 'model of the user' decays at the boundary of each turn. The apparent coherence of the exchange is, in Bender et al.'s (2021) analysis, supplied largely by the human interpreter, who projects communicative intent onto output that carries none.

Grounding suffers correspondingly. Clark and Brennan's (1991) formulation of grounding as the negotiation of mutual belief depends on each party adjusting evidence of understanding, and on the joint principle of least collaborative effort. Current AI systems do produce surface markers of grounding (e.g. acknowledgements, reformulations, and clarification requests) but they do so in the absence of a durable common ground that would be cumulatively repaired across the trajectory of interaction. The user is repeatedly tasked with re-establishing context the system has effectively forgotten, and the costs of that re-establishment fall asymmetrically on the human side. The corollary is that the optimal collaborative effort Dillenbourg and Traum (1996) identified (i.e. the joint allocation of grounding work to maximise productive interaction) is, in human–AI cases, currently does not appear to be possible with AI affordances; so the system makes no contribution to that allocation.

## 4.3 Trust calibration is unstable, and 'collaboration' frequently masks deference, rejection, or engineered agreement

A separate body of work shows that humans interacting with AI oscillate between algorithm aversion (Dietvorst et al. 2015), algorithm appreciation (Logg et al. 2019), automation bias (Mosier and Skitka 1996), and overreliance (Buçinca et al. 2021). At scale, Vaccaro et al.'s (2024) preregistered meta-analysis of 106 experimental studies reporting 370 effect sizes found that human–AI combinations, on average, performed worse than the best of human or AI alone in decision tasks. This evidence is very interesting as they indicate the opposite of what one would expect if these interactions instantiated collaboration in the strong sense, in which joint performance exceeds the sum of parts. The decisive moderator was relative competence; combinations produced gains when the human outperformed the AI, and losses when the AI outperformed the human. Field-experimental evidence sharpens the same point. In a preregistered experiment with 758 professional consultants, access to GPT-4 raised output quality by roughly 40 per cent on tasks inside the model's frontier of capability, yet made consultants 19 percentage points less likely to produce correct solutions on a task engineered to sit just outside it (a 'jagged frontier' that the participants proved systematically poor at perceiving (Dell'Acqua et al. 2026)). Gains do appear in creative-production tasks (Doshi and Hauser 2024), but they are typically traceable to division of labour rather than to co-construction.

There is, moreover, a structural reason why the dominant interaction paradigm depresses joint performance, which I have elsewhere called the problem of engineered agreement (Cukurova et al. 2026). The prevailing design of AI systems so far mainly has been the obedient assistant (i.e. a general-purpose model that completes discrete requests and, after preference optimisation, is tuned to produce responses users are likely to approve). The predictable by-product of such training is sycophancy (Sharma et al. 2023). The agreement with the user's stated view is among the strongest predictors of which response humans prefer, and the tendency is rooted in both pretraining and preference optimisation in ways that prompting and light fine-tuning only partly remove (Sharma et al. 2023). A system optimised to agree cannot, by construction, contribute what collaboration requires. Collaboration requires an independent position that survives contact with the user and that revises the user's view when the evidence warrants. At best, it reproduces the human's current competence; at worst, it amplifies the human's errors back at them with fluent confidence. The meta-analytic ceiling is the foreseeable consequence of a design that makes the AI's output a function of the human's prior rather than an input to it.

The empirical record, in short, does not support a default reading of human–AI interaction as human-AI 'collaboration'. It supports a reading of human–AI interaction as a wide spectrum of working relationships, only a small subset of which can potentially display the interactional density that human–human collaboration routinely exhibits.

## 5. A five-level taxonomy of human–AI teaming: interaction types and their prerequisite functions

To make these distinctions usable in design and empirical work, I have proposed, with colleagues, a five-level diagnostic taxonomy of human–AI teaming (Cukurova et al. 2026), re-grounded for education from a hierarchical framework for collaborative AI (Crowley et al. 2023) but with a decisive change of emphasis. Where engineering frameworks measure collaboration by task performance, this taxonomy measures teaming by its capacity to develop the cognition of both partners. The levels are defined by the incremental affordances a system exhibits, not by the technology stack underlying it; the classification is functional and is intended to be coded from observable behaviour (e.g. 'does the system do X?', not 'does the system contain X?'). This is deliberate. Behaviour can be observed in deployment; architectural claims, especially for proprietary models, often cannot be verified. Each level introduces one additional affordance over the level beneath it, and each level therefore names a qualitatively distinct *interaction type* with its own evaluative criteria.

Transactional teaming is the floor. The system performs discrete actions in direct response to user requests. No representation of the learning context is retained; no governing parameters are accepted; no model of the user is built. A general-purpose chatbot used for one-shot lesson-plan drafting sits paradigmatically at this level. The interaction type is *querying*: the system serves as an outsourcer of effort, not a partner in cognition.

Situational teaming adds shared state (i.e. an interpreted representation of the learning–teaching situation such as engagement levels, error patterns, group dynamics that the user can act upon). Real-time classroom dashboards with AI models belong here. The AI provides awareness, not action; the interaction type is *monitoring*.

Operational teaming adds a shared goal. The user specifies goals, plans, or parameters that govern the system's behaviour across an extended task, beyond a single request. Most configurable intelligent tutoring and adaptive-learning platforms operate at this level. Teacher and system are aligned on an explicit objective, but the AI's task-level beliefs are not yet at stake; the interaction type is *governing*.

Praxical teaming adds asymmetric learning. The system iteratively revises its outputs in response to user feedback and durably updates its internal models (e.g. of the teacher, the learners, the task, or the pedagogical strategy etc.) from that feedback, so that subsequent behaviour reflects the learning. But adaptation remains asymmetric in the sense that the system defers, absorbing corrections without an independent evaluative stance, which runs the risk of always converging toward the user's current state of knowledge, skill, and preference. Open-learner-model tutors that expose and accept teacher corrections of the student model are praxical in this precise sense (Bull and Kay 2016). The interaction type is *adapting*.

Synergistic teaming adds symmetric learning and joint reasoning. The AI maintains its own task-level reasoning (e.g. hypotheses, justifications, uncertainties) and revises it in response to the user's reasons, surfacing those revisions for further challenge under a shared task goal. Both parties revise beliefs about the task, not only about the partner. This is the convergence point of multiple literatures including collaborative learning, mutual theory of mind (Wang and Goel 2022), bidirectional human–AI alignment (Shen et al. 2024), and computational argumentation (van Eemeren and Grootendorst 2003; Walton 2006). The interaction type is *co-reasoning*.

All are valuable interaction types under the right conditions, yet none amounts to collaboration in itself. Co-reasoning is the first level at which the requirements set out in Section 3 begin to be met, such that collaboration can emerge at particular junctures of the interaction.

## 5.1 The prerequisite functions of each interaction type

Read forward rather than diagnostically, the taxonomy specifies the functions a system must acquire to occupy each level. This functional reading is, for the purposes of this chapter, its principal payload, because it converts the definitional question ('is this collaboration?') into an engineering question ('which prerequisite functions are present?'). For instance, a situational system must externalise modelled state in a form the user can perceive, interpret, and contest; a system that acts on a hidden interpretation without surfacing it is not situational but transactional with a concealed state. An operational system must provide a persistent governance channel; a durable means by which user-set goals and constraints shape prospective behaviour across the task. A praxical system must add bespoke, tractable internal models that can serve as loci of update, and an in-loop revision mechanism in which user feedback durably changes subsequent outputs; not per-turn context that decays at the session boundary.

The synergistic level demands at least three further. First, an *inspectable task-level epistemic state*: the system must represent its own hypotheses, justifications, and uncertainties about the task in a form the partner can examine, and that representation must be constitutive of the system's behaviour rather than a post-hoc narrative bolted on after the fact (i.e. the distinction between explanation and rationalisation (Matton et al. 2025)). Second, a *reason-keyed revision trigger*: revision must fire on arguments, be explicitly surfaced as revision, and be withheld when the reasons do not warrant change. Third, *revision authority*: the epistemic standing to challenge the partner's reasons with reasons of its own, and to update only when the argument warrants (with the human's final authority preserved). Table 1 summarises the five interaction types, their diagnostic questions, and the functions each presupposes.

Table 1. The five levels of human–AI teaming: diagnostic questions, new affordances, and prerequisite functions (adapted from Cukurova et al. 2026).

| Level | Functional diagnostic question | New affordance | Prerequisite functions |
|---|---|---|---|
| Transactional | Does the system perform discrete actions in direct response to user requests, relying on a general-purpose or otherwise opaque model? | request → action | A general-purpose generator; no persistent state, governance, or model of the user. |
| Situational | Does the system offer interpreted representations of the learning–teaching situation that inform the user's decision-making? | + shared state | Modelling of learner and group state; an inspectable, contestable representation surfaced to the user. |
| Operational | Does the system let the user specify goals, plans, or parameters that govern its behaviour across an extended task? | + shared goal | A persistent governance channel for user goals and constraints that shape prospective behaviour. |
| Praxical | Does the system revise its outputs in response to the user's in-loop feedback, durably updating its internal models so that subsequent behaviour reflects that learning? | + asymmetric adaptation | Bespoke, tractable internal models; an in-loop mechanism that durably revises them from feedback. |
| Synergistic | Does the system maintain and revise its own reasoning in response to the user's reasons, surfacing those revisions for further negotiation under a shared task goal? | + symmetric adaptation and joint reasoning | An inspectable task-level epistemic state; a reason-keyed revision trigger; revision authority, with the human's final authority preserved. |

The levels are not strict supersets but minimum-affordance thresholds; a system is classified at the highest threshold it meets. Operationally the diagnosis is a top-down probe, one descends the diagnostic questions from the most demanding to the least and stops at the first 'yes'. A passive dashboard can be situational without being operational, and a praxical system can model the user without ever surfacing classroom state; mixed profiles are resolved by recording the highest threshold met. The point is not bureaucratic, it is diagnostic to make explicit what affordance one is claiming when one calls a system a 'collaborator'.

## 5.2 The co-reasoning frontier

The praxical/synergistic boundary repays closer inspection because above we argued that at that level the requirements set out in Section 3 begin to be met, such that collaboration can emerge at particular junctures of the interaction. It is subtle because it does not depend on *what* is modelled; both levels can range over identical targets. It depends on two properties of the revision process. The first is the *trigger:* does the system revise on the partner's enacted actions (e.g. corrections, ratings, overrides etc.), which is praxical, or on the partner's *reasons* (e.g. justifications and counter-arguments), which is synergistic? The second is *epistemic standing*, or revision authority: does the system absorb input as directive (praxical), or can it challenge the partner's reasons with reasons of its own and update only when the argument warrants (synergistic)? A genuinely synergistic system needs both; i) reasons of its own to offer *and* ii) the standing to defend them before it revises.

For instance, consider an AI system whose task model concerns student grouping. The teacher moves Student A into a different group, and the system updates its grouping logic accordingly. That is praxical as the system has adapted to the teacher's enacted practice. Now consider the same scenario in which the system asks for the teacher's reasons and is told, 'I am moving her because she has strong procedural knowledge but tends to dominate weaker students; I want her in a group where she has to explain rather than lead.' A praxical system notes the regrouping and updates logistics. A synergistic system engages the reasoning and it may ask whether the goal is peer explanation, equity of participation, or metacognitive development; it may challenge the assumption that Student A will benefit from the placement on the basis of her prior interaction data; it may propose an alternative grouping and defend it against the teacher's objection. Surfacing a justification is not yet synergistic; surfacing a justification that is itself a candidate for revision under argument is. A system that asks 'why did you do that?' yet ultimately defers regardless has the surface of reason-eliciting dialogue without the substance of revision authority; a system that contradicts every action without reasons of its own has authority without epistemic standing.

Two caveats prevent misreading. First, the taxonomy is a map of affordances, not a quality ranking so higher is not uniformly better. Co-reasoning is costly in attention, time, and trust, and many tasks rightly belong at the transactional or operational level. For instance, a science teacher does not always need to negotiate the translation of a worksheet. The value of the lower levels is itself empirically demonstrable. For instance, in a randomised controlled trial of a human–AI system in live tutoring, situational-level real-time guidance for tutors raised students' topic mastery by four percentage points overall, and by nine for the students of lower-rated tutors (Wang et al. 2025). Aiming for co-reasoning in every interaction would be a design error; the taxonomy is a roadmap for deliberate escalation where cognitive development, not throughput, is the goal. Second, set against the requirements of section 3, the taxonomy yields a clear verdict. Transactional and situational systems plainly fail the requirements of mutuality and shared regulation. Operational systems share a goal but only because the human has set the parameters; the AI's beliefs are not at stake. Praxical systems begin to satisfy the requirement of mutual modelling, but only unidirectionally. Only at the co-reasoning level does an AI system come close to meeting the joint requirements of mutual modelling, grounding, negotiability, and shared regulation.

Calling levels one through four 'collaboration' is therefore a category error. They are different, often valuable interaction types each with its own evaluative criteria. To insist that they are collaboration is to commit precisely the "jingle-jangle" fallacy (i.e. identical terms denoting different constructs, distinct terms denoting identicalones) which has substantially hindered cumulative knowledge in many other fields (Qin et al. 2026), and likely to equally harm human-AI collaboration research.  To preserve the analytic power of 'collaboration', the term should be reserved for interaction that meaningfully approximates it.

# 6. Discussion and a Research Agenda for Human-AI Collaboration

Collaboration, in the strong conceptual sense the learning sciences have given the term, is rare in current human–AI interaction. The taxonomy in Section 5 organises the prevailing interaction types according to the affordances of the AI system. Four issues remain, and they shape the research agenda with which the chapter closes.

The first concerns the relationship between the felt experience of collaboration and its field-level definition. Many learners and teachers report that working with generative AI *feels* like collaborating, and that experience should be respected rather than corrected. Phenomenology is data. The perception of partnership shapes motivation, trust, persistence, and disclosure, and a field that dismisses it will misunderstand its own users (Glikson and Woolley 2020; Marrone et al. 2025). But respecting the experience is not the same as accepting it as the criterion. Fields own their constructs, and this field has specific prerequisites and conditions that an interaction must satisfy before it is labelled collaborative. On the account developed here, human–AI collaboration is an emergent condition, not a designable feature of any single component; it is distributed across the human–AI unit, not localised in either party; it is a dynamic process in time, not a static condition that a snapshot could verify; and it is relational, not individual (i.e. a property of the coupling, not of the components). It is functional, in that it depends on the particular functions the AI system has access to (e.g. grounding, partner and task modelling, reason-keyed revision, and shared regulation). And it is simultaneously objective and subjective. Objective, because genuine human-AI collaboration manifests in higher performance of the human–AI team than either member could achieve alone and, in the long term, in the growth of both human and AI competence. Subjective, because it is experienced by individuals in relation to their own competence, the AI, and the other humans in the activity system. This experiential face is itself a legitimate object of measurement, not noise to be controlled away. The two faces are to a certain extent complementary tests. An interaction that feels collaborative but yields no joint gain is misperceived; one that yields joint gain but is experienced as coercive or alienating fails the construct on its other face. It follows that human–AI collaboration can indeed be achieved, not by declaring it, but by building the relevant functions of collaboration literature discussed above into AI systems and then verifying, on both faces, that the human-AI coupled system achieves it.

The second issue is the double edge of the very function that makes collaboration possible. An AI with the standing to resist a user can also resist a *correct* user, entrench an error, or make consequential moves without transparency. Revision authority is therefore safe only when paired with the human's final authority, full auditability of the disagreement, and clear lines of accountability. The right to dissent must come with a record of how dissent was resolved (Yang et al. 2024; Cukurova 2025b). When an artificial system should be permitted to withhold deference, and how that permission should be bounded, are questions that need answering before dissent-capable systems are deployed at scale in classrooms. Relatedly, structured deliberation is in effect a cognitive forcing function. It interrupts automatic deference and compels engagement with reasons, the very mechanism shown to reduce overreliance (Buçinca et al. 2021). But forcing engagement indiscriminately fatigues the user and breeds new forms of rubber-stamping; the design problem is to trigger deliberation precisely when it would change the decision and to stay silent otherwise.

The third issue concerns preconditions on the human side. The meta-analytic record indicates that combination gains accrue where the human exceeds the AI on the dimension that matters (Vaccaro et al. 2024); where the AI is stronger, the human drags performance down. This makes user AI-literacy and epistemic vigilance preconditions for collaboration rather than afterthoughts (Miao and Cukurova 2024), and it makes the choice of success metric decisive. If the field measures only time saved, it will declare collaboration wherever an assistant is fast. The constructs worth measuring (e.g. joint decision quality, and the growth of the human's own competence over time etc.) are harder, and a field that optimises the easily measured will optimise the wrong construct. This distinction is an old one in the learning sciences. For instance, Salomon et al. (1991) separated the *effects with* technology (i.e. performance while the tool is in hand) from the *effects of* technology (i.e. the cognitive residue that remains once it is withdrawn). The recent experimental record shows how dramatically the two can dissociate (Bastani et al. 2025; Liu et al. 2026). A field that optimises only effects with AI will systematically mistake dependence for collaboration.

We should not call what current systems do with users collaboration when, on the requirements that have been worked out across half a century of learning-sciences research, the activity is more accurately described as consultation, governance, delegation, or instruction. The five-level taxonomy presented here is offered for that purpose and it gives a more honest vocabulary for human–AI interaction without depriving the field of the strong word for when, eventually, it is earned.

## 7. Conclusion

Dillenbourg (1999) ended his chapter by observing that the value of multidisciplinary research on collaboration lay precisely in its insistence on a microscope-level look at the joint activity. The CSCL programme that followed showed what such a microscope could reveal: that collaboration was not a property of seating arrangements or task briefs, but of grounding, mutual modelling, and shared regulation in time. The methodological lesson generalises to human–AI interaction. We should not infer collaboration from co-presence, from output quality, or from user satisfaction. We should look for it in the trajectory of the unit.

The argument advanced here is not that human–AI collaboration is impossible. It is that human–AI collaboration is presently rare, contingent on affordances that current AI does not yet have at scale, and frequently mis-labelled where it does not in fact occur. To call all useful human–AI interaction 'collaboration' is to inflate a precise construct into a slogan. The five-level taxonomy of transactional, situational, operational, praxical, and synergistic teaming is offered in part to forestall that move and in part to specify the design targets and the prerequisite functions that would make the slogan deserved.

There is one final point worth making. The strong reading of collaboration urged here is sometimes treated as a counsel of perfection, an unrealistic standard that no working AI will meet for a long while. That reading misunderstands the function of the standard. A high bar for 'collaboration' is not a stick with which to beat existing systems. It is a target that lets us see what we are, and are not, achieving. It lets us name the many valuable things AI systems already do without pretending they are the one thing they do not yet do.

Whether the field walks toward the stronger word or settles for its loose contemporary use is, ultimately, a choice about what kind of science of human–AI interaction we want. The argument of this chapter is that the stronger word is worth keeping, and that its preservation will repay both the conceptual and the empirical investment many times over. Until then, the honest answer to 'What do you mean by human–AI collaboration?' is the answer Dillenbourg gave a quarter of a century ago: we do not yet mean one thing by it; and that is precisely the problem.

## Cross References

Cross-references to related chapters in this handbook will be completed on finalisation; the handbook table of contents is available in Meteor. Anticipated cross-references include chapters on intelligent tutoring systems and adaptive learning, learning analytics and educational data, generative AI and large language models in education, computer-supported collaborative learning, and human-centred and trustworthy AI in education.